\documentclass[superscriptaddress,twocolumn,aps,prb,showpacs]{revtex4-1}
\usepackage{graphicx}
\usepackage{amsbsy,amssymb,amsmath,bm,ulem}

\usepackage{color}

\newcommand{\comment}[1]{}

\renewcommand{\Re}{\mathop{\mathrm{Re}}}
\renewcommand{\Im}{\mathop{\mathrm{Im}}}

\graphicspath{{../fig/}}

\begin{document}

\title{Ratchet effects for paramagnetic beads above striped ferrite-garnet films}
\author{J. I. Vestg{\aa}rden}
\author{T. H. Johansen}
\affiliation{Department of Physics, University of Oslo, P. O. Box
1048 Blindern, 0316 Oslo, Norway}

\begin{abstract}
We calculate the motion of a small paramagnetic bead which is
manipulated by the stripe domain pattern of a ferrite-garnet film.  A
model for the bead's motion in a liquid above the film is developed
and used to look for ratchet solutions, where the bead acquires net
coherent motion in one direction when the external field is modulated
periodically.  We consider three cases. 
First, the ratchet, where the
beads all go in the same direction.  
Second, the height dependent ratchet, where
beads at different heights go in opposite direction.  This case can be
used to separate beads of different sizes, as considered in
J. Phys. Chem. B 112, 3833 (2008). 
Third, we describe how the
separation threshold can be tuned by changing the amplitude of the
applied field.  
Finally, we describe a pseudo ratchet, where
the external modulation is not periodic and the ratchet changes
direction periodically.  
\comment{poeng: siden effekten er robust mhhp $\beta$
vil den ogs\aa~funke med biologisk last.}  \comment{poeng: Dhar07
regner p\aa~rathet for buede domener (labyrinter)}
\end{abstract}

\maketitle

\section{Introduction}
Functionalized micrometer-sized beads have for a long time been used
for medical and biological applications where a typical application is
to attach biologically active molecules to the beads and use the beads
as carriers.  However, the setup for such applications is usually
limited to bulk manipulation where the beads are contained in an
aqueous solution, and such setups do not offer precise positioning of
individual beads. Recently this situation has improved and several
devices demonstrating controlled manipulation of a small number of
beads have been realized, for example by using a thin zig-zag electric
wire\cite{vieira09} or by a hybrid device magnetic bead separator
device using current wires and magnetic fields combined with
microfluidic channels.\cite{smistrup07}

One promising way to realize devices that manipulate individual beads
is to use domain patterns in ferrite garnet films.  (For a review,
see. Ref.~\onlinecite{tierno09}) Because the paramagnetic beads get
strongly pinned to the domain walls the influence of thermal noise and
other perturbations is small. Also, the beads can easily be moved from
the domain walls by an external field interacting both with the beads
and the domain pattern.  An additional level of control can be
archived by monitoring the domain patterns by exploiting the larger
Faraday rotation of ferrite garnet films.  All these features where
demonstrated in Ref.~\onlinecite{tierno07} in which the stripe pattern
of a ferrite garnet film in an harmonic applied field created a
magnetic ratchet effect, where beads jumped from domain wall to domain
wall in just one direction. In the same system, with a more complex
applied field, Ref.~\onlinecite{tierno08} demonstrated an even more
surprising effect: beads of different sizes go in opposite directions.
The motion in this double ratchet, as as in the original rathet, was
synchronized so beads jump only one wavelength of the stripe pattern
per period of the external field.

In this work we model the the system of Refs.~\onlinecite{tierno07}
and ~\onlinecite{tierno08} and discuss the cause of the effects and
tuning of the devices.

\comment{put in proper place} 

\begin{figure}
  \centering
  \includegraphics[width=\columnwidth]{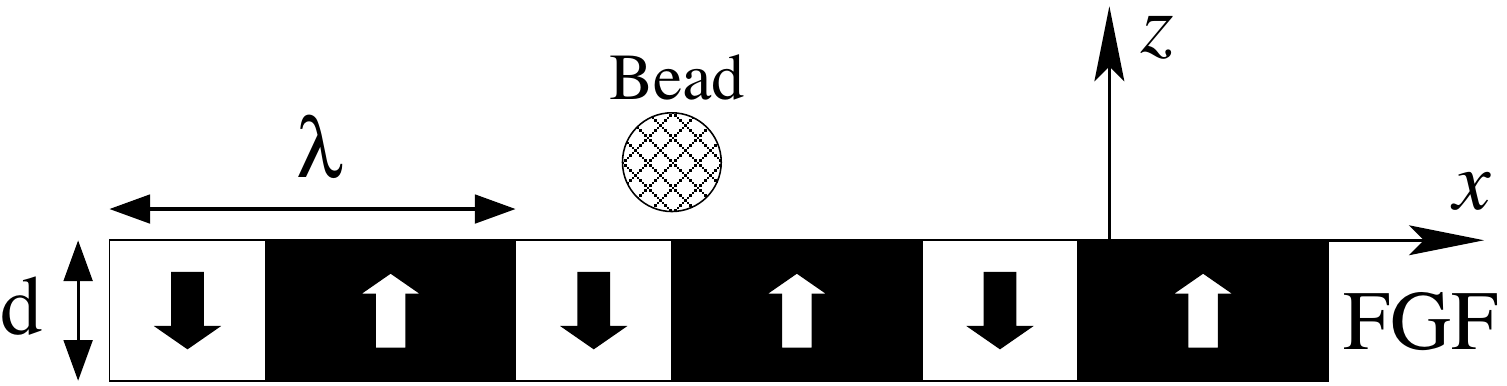}
  \caption{A side-view of the stripe domain pattern of a ferrite garnet film. The film 
    has thickness $d$ and pattern wavelength is $\lambda$. A paramagnetic bead with radius $a$
    is pinned to the domain wall.}
  \label{fig:sample}
\end{figure}

\section{Model}
Consider a paramagnetic bead dispensed in a liquid above the surface
of the film.  The bead motion is mainly determined by the hydrodynamic
force $F^h_x$ and magnetic force $F^m_x$.  Then the equation of motion
is
\begin{equation}
  F^h_x+F^m_x=0
  \label{dynamics-x}
  .
\end{equation}
This overdamped motion is a reasonable approximation for slow dynamics
and it is also reasonable to neglect other forces, provided the beads
avoid direct contact with the surface.  The hydrodynamic drag of
slowly moving beads dispensed in a liquid is quantified by Stoke's law
\begin{equation}
  F_x^h = 6\pi fa\eta~\dot x
  \label{stokes-law}
  ,
\end{equation}
where $a$ is bead radius, $\eta$ is the dynamic viscosity of water and
$f$ is a correction factor due to the presence of the film,
$f=f(z)\geq 1$. In experiments the film is typically electrostatically
charged to prevent sticking, so that the beads leviate a distance a
few nanometer above the film, yielding $f<3$.\cite{helseth06}

The magnetic force on the paramagnetic bead with volume $V$ and
magnetic susceptibility $\chi$ is
\begin{equation}
   F_x^m = -\frac{\partial U}{\partial x}
  \label{force_x}
  .
\end{equation}
with potential 
\begin{equation}
  U = -\frac{1}{2}\mu V\chi H^2
  \label{potential-u}
\end{equation}
where $\mu \approx \mu_0= 1.26\times 10^{-6}$~Tm/A is the permeability of water,

Consider a magnetic film with stripe domains. Above the surface of the film, 
the total magnetic field is
\begin{equation}
  \mathbf H(\mathbf r, t) = \mathbf H^a(t) + \mathbf H^f(\mathbf r,t)
\end{equation}
where $\mathbf H^a$ is applied field and $\mathbf H^f$ is the inhomogeneous 
field from the stripe pattern of the film. The applied field 
is periodic, with 
\begin{equation}
  \label{def-Ha}
  \begin{split}
    &H_z^a=H_0\sin(2\pi\nu t) \\ 
    &H_x^a = sH_0\sin(g2\pi\nu t)  
  \end{split}
\end{equation}
where $g$ is an integer and $s$ is an arbitrary number.

\begin{figure}[t]
  \centering
  \includegraphics[width = \columnwidth]{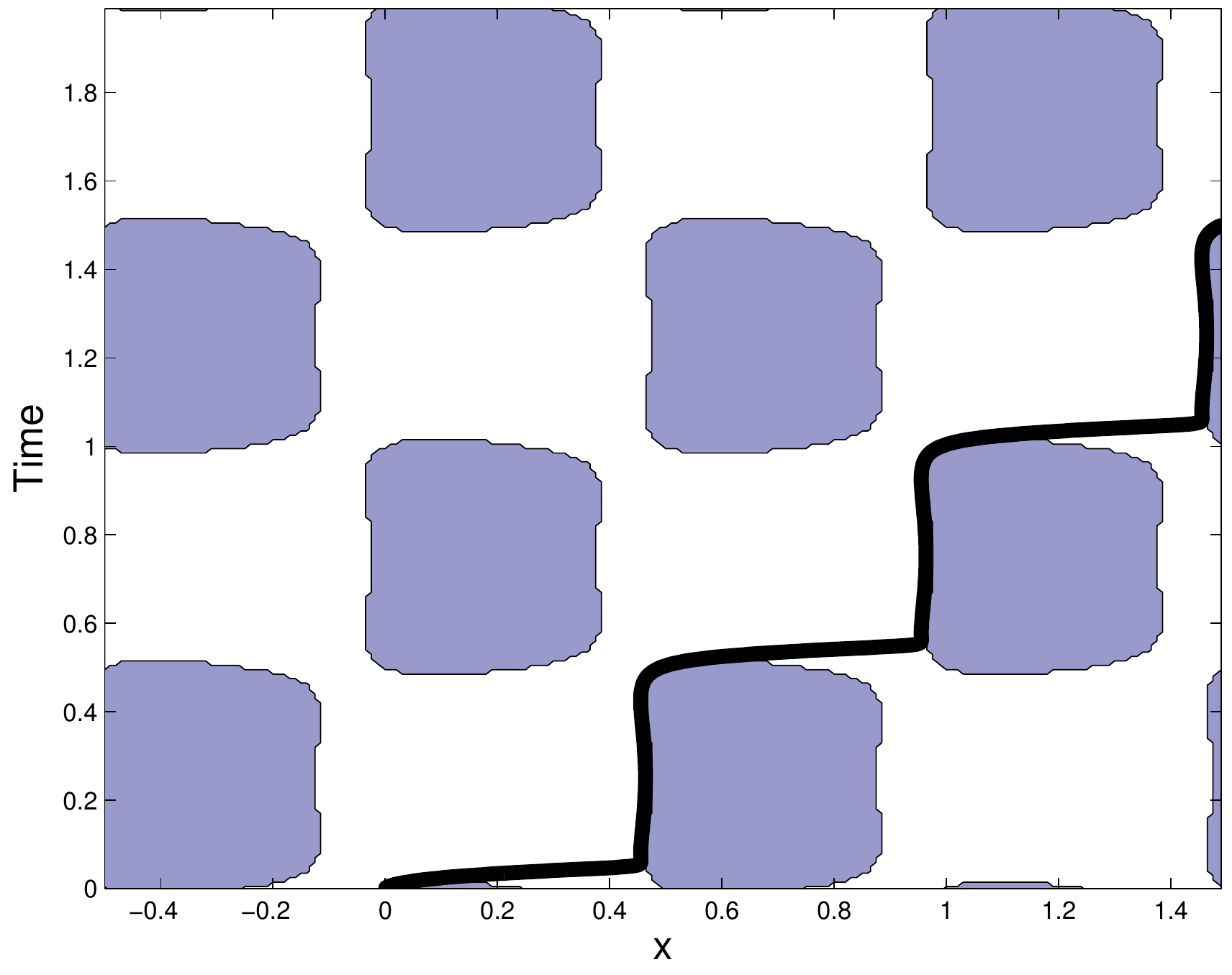} 
  \caption{
    The ratchet effect. The horizontal axis is the position $x$ and the vertical axis is time $t$. 
    The particle path is plotted as a solid line 
    above the landscape formed by the sign of $F_x^m$. Blue means $F_x^m<0$ and 
    white is $F_x^m<0$
    \label{fig:ratchet}
  }
\end{figure}

\begin{figure}[t]
  \centering
  \includegraphics[width = \columnwidth]{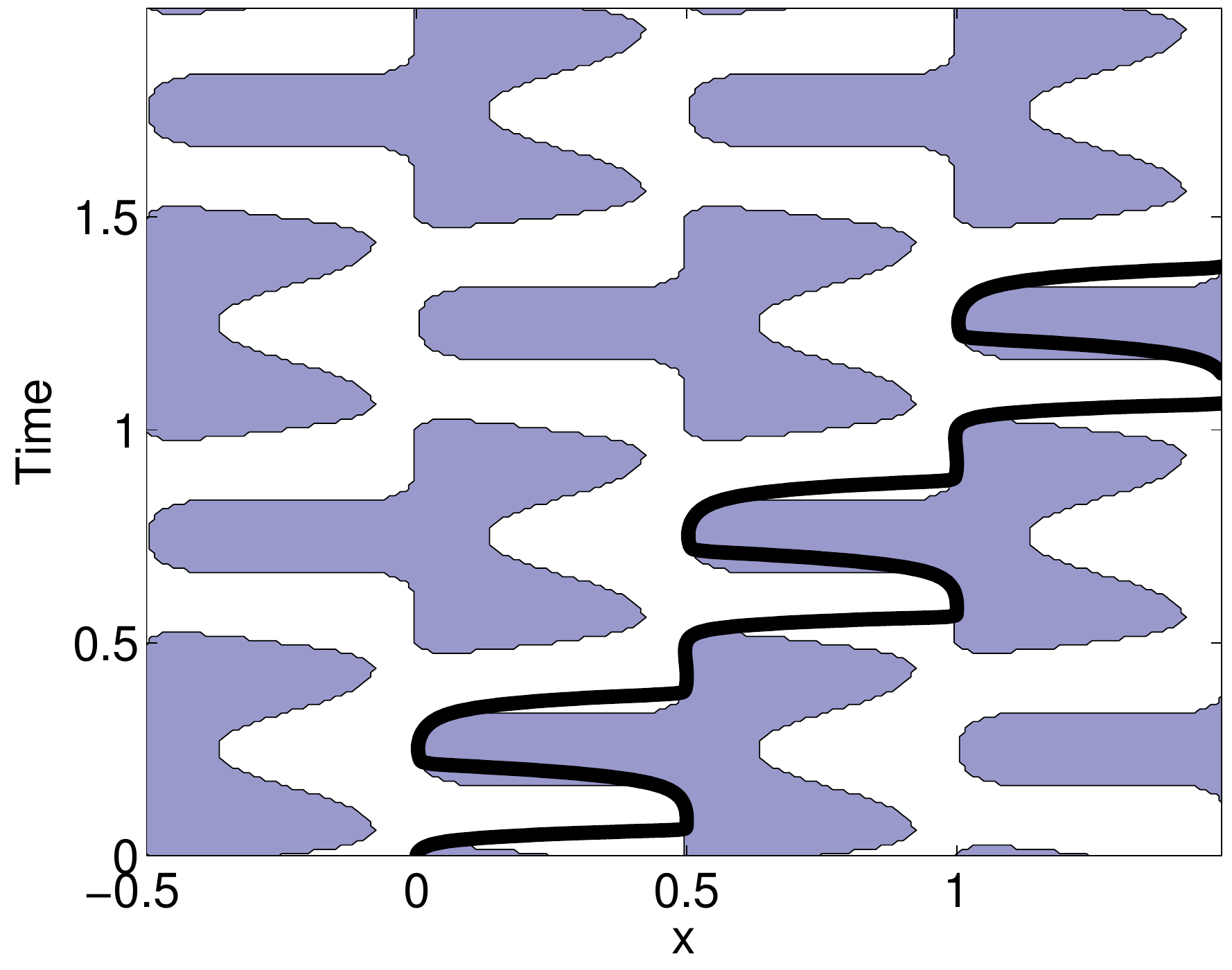} \\
  \includegraphics[width = \columnwidth]{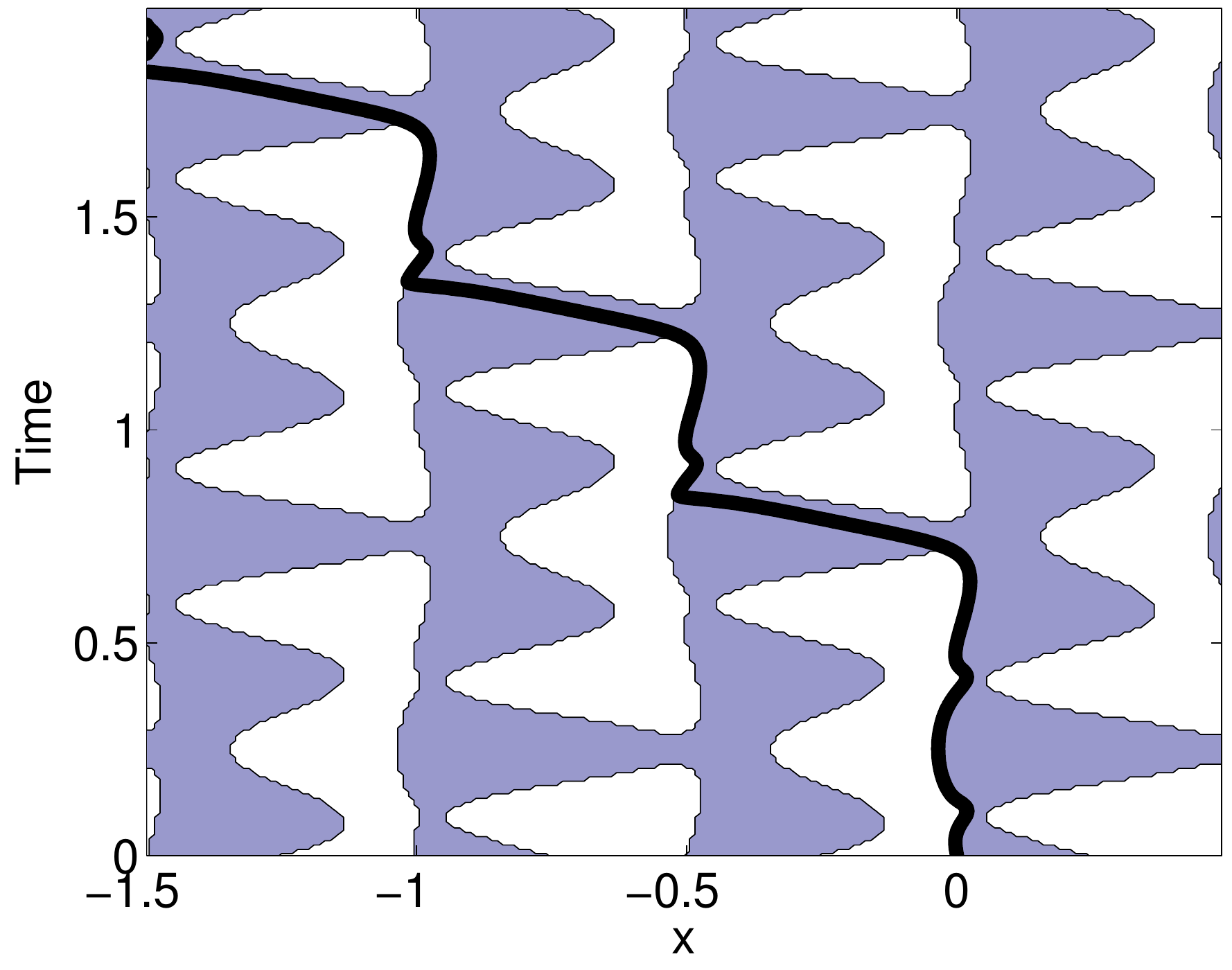}
  \caption{Separation effect, in which beads at different heights move in opposite directions. 
    The bead's trajectory is plotted above the sign of the interaction force. Top:  
    $z=0.1\lambda$, bottom: $z=0.05\lambda$. All other parameters are equal: $H_0=0.18 M_s$ 
    and $\Delta_0=0.05$.
    \label{fig:separation}
  }
\end{figure}

For a film with stripe domains and magnetization $M_s$, 
the magnetic field above the film can be expressed as 
\begin{equation}
\begin{split}
  &H_x^{f}=\frac{M_s}{\pi}\Re\psi \\
  &H_z^{f}=\frac{M_s}{\pi}\Im\psi 
\end{split}
  \label{Bx-and-Bz}
\end{equation}
where the complex auxiliary function $\psi$ is 
\begin{equation}
\begin{split}
  \psi=
  & \log\left( 
  \frac{1-e^{-2\pi(z+i(x+\Delta))/\lambda}}
       {1+e^{-2\pi(z-i(x-\Delta))/\lambda}} 
  \right) \\
   + 
   & \log\left( 
  \frac{1+e^{-2\pi(z+d+i(x+\Delta))/\lambda}}
       {1-e^{-2\pi(z+d+i(x-\Delta))/\lambda}} 
  \right)
  .
\end{split}
\label{psi}
\end{equation}
where $d$ is film thickness, $\lambda$ is the wavelength of the stripe
pattern and $\Delta$ is the domain wall displacement.  The second term
in Eq.~\eqref{psi} comes from the bottom surface of the film and it
can be neglected when beads are close to the film, i.e., when $z\ll
d$.  The domain wall displacements depends on the applied field, and
for for small and medium amplitudes we assume it to be proportional to
$H_z^a$, i.e.
\begin{equation}
  \Delta(t)= \Delta_0\frac{H_z^a(t)}{H_0}
  .
\end{equation}

\section{Results}
When the external field is modulated there will be response in the
width of the strip domains and the paramagnetic bead will move
according to Eq.~\eqref{dynamics-x}.  We will now consider several
solutions: the ratchet, the pseudo ratchet and the size dependent
ratchet.

\emph{Ratchet -} 
An example of coherent dynamics is the 
ratchet. In the ratchet the paramagnetic beads gain a net coherent motion 
in one direction, despite a periodic modulation of the external field.
The bead's motion is plotted in Fig.~\ref{fig:ratchet}. The external field is in this setup constantly
at angle 45$^\circ$ corresponding to $s=1$ and $g=1$ in Eq.~\eqref{def-Ha}. The motion 
of the will nonetheless acquire a coherent motion to the right, as seen in Fig.~\ref{fig:ratchet}.

\emph{Separation -} Complex modulation of $\mathbf H_a$
opens the possibility even more strange ratchet solutions of
Eq.~\eqref{dynamics-x}, for example ratchets that go in opposite
directions at different heights. This effect was exploited in
Ref.~\onlinecite{tierno08} to yield separation of beads of different
sizes.  The height dependent ratchet appears for $g=3$ in
Eq.~\eqref{def-Ha} and the results trajectories for beads at heights
$z=0.1\lambda$ and $z=0.05\lambda$ are plotted in
Fig.~\ref{fig:separation}.  There is a clear difference: the large
bead jumps three times during a half-period and acquires a net motion
to the right while the small bead jumps only once and acquires a net
motion to the left, all in perfect agreement with
Ref.~\onlinecite{tierno08}.  The two trajectories are entirely as
expected from the sign of $F_x$ which are plotted in the
background. However, we notice that the beads do not follow the edge
of the signs exactly, because of the finite viscosity. Only in the
limit $\eta\to 0$, the beads would follow the edges exactly.

\begin{figure}[t]
  \centering
  \includegraphics[width = \columnwidth]{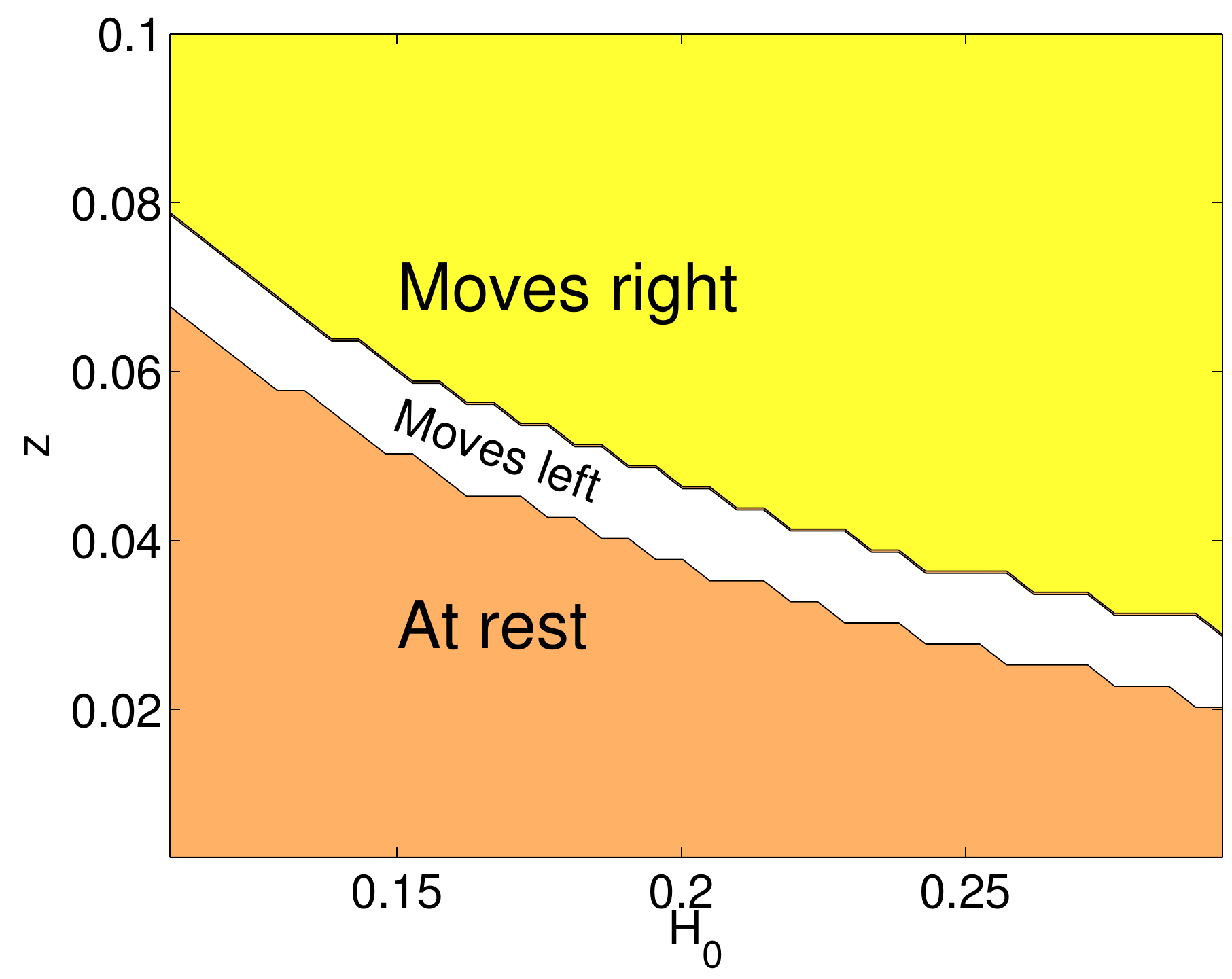} \
  \caption{
    Tuning of the separation height by changing 
    the strength of the applied field $H_0$ from Eq.~\eqref{def-Ha}.
    The part with inverted ratchet direction is seen as a white line.
    \label{fig:phase-diagram-H0-z}
  }
\end{figure}

\begin{figure}[t]
  \centering
  \includegraphics[width = \columnwidth]{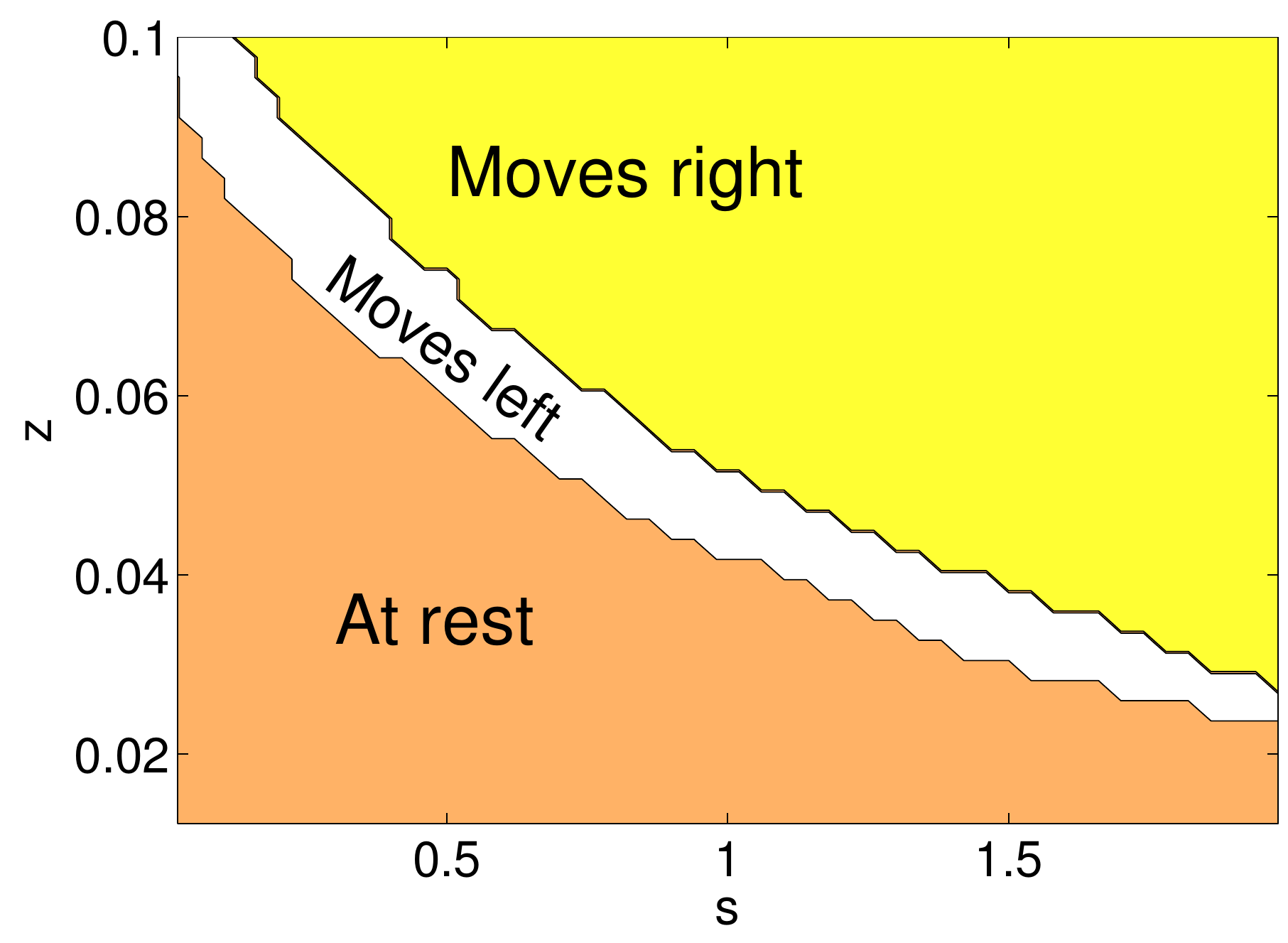} 
  \caption{
    Tuning of the separation height by the amplitude asymmetry of the 
    applied field $s$ from Eq.~\eqref{def-Ha}. $H_0=0.18M_s$.
    The part with inverted ratchet direction is seen as a white line.
    \label{fig:phase-diagram-s-z}
  }
\end{figure}

\emph{Tuning of separation threshold -} The height dependent ratchet
will only be useful as a separation device if the separation threshold
$z_\text{th}$ is possible to tune. The only realistic parameter to
change is $\mathbf H_a$, but since the film reacts to $H^a_z$ the most
robust approach is to keep $H^a_z$ fixed and just change $H^a_x$.  The
effect of changing $s$ is seen in Fig.~\ref{fig:phase-diagram-s-z}
which is plotted in the style of phase diagram, distinguishing tree
outcomes: motion to the left, to the right and at rest.

\begin{figure}[b]
  \centering
  \includegraphics[width = \columnwidth]{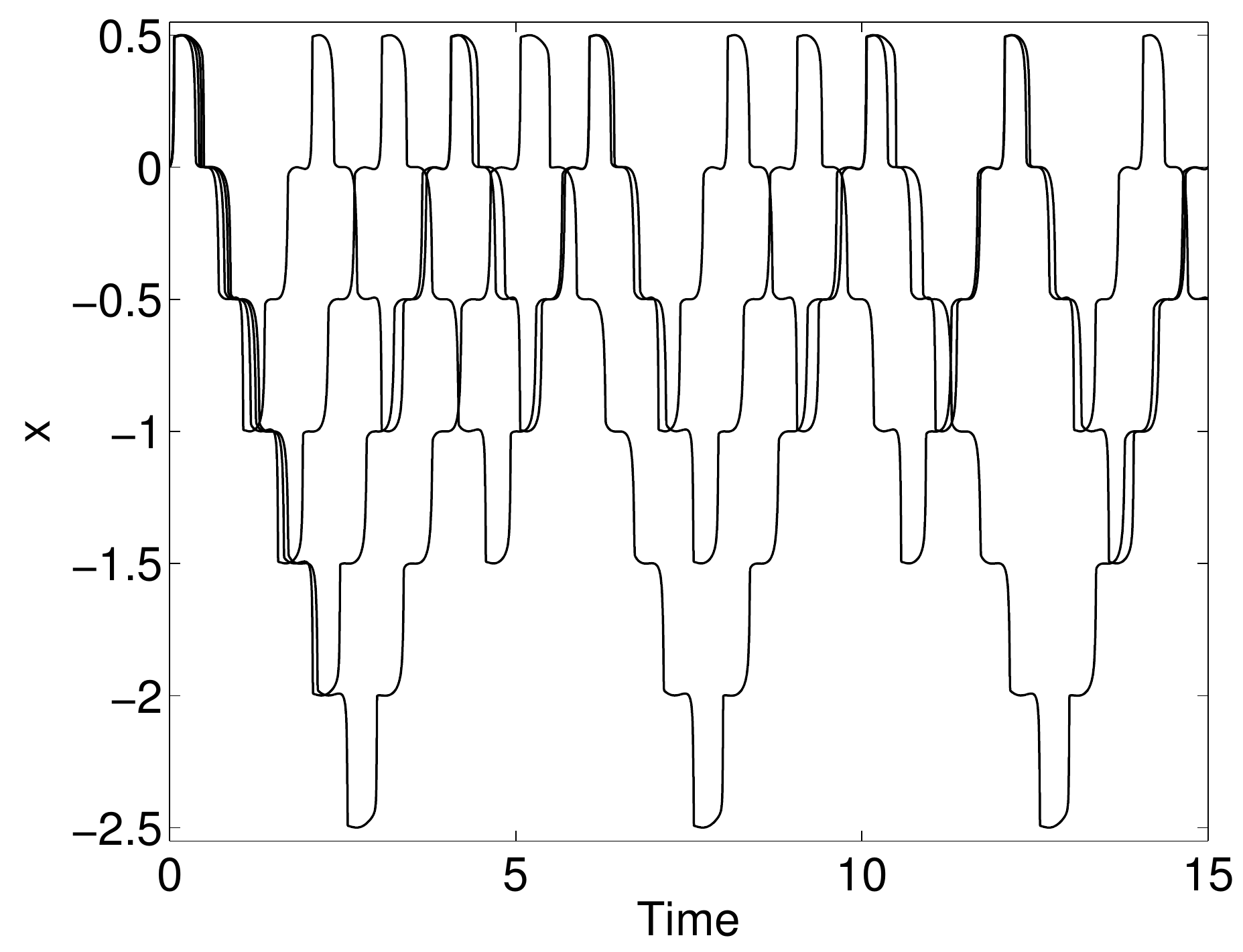} 
  \caption{
    Pseudo ratchet, in which the ratchet direction changes after a $N=2,3,4$ 
    and 5 periods.    
    \label{fig:pseudo-ratchet}
  }
\end{figure}

\emph{Pseudo ratchet -}
The ratchet effect is a consequence of phase coherence between the
$H^a_x$ and $H^a_z$ and the ratchet fails when there is a drift in
$H^a_x$, i.e., when $g$ from Eq.~\eqref{def-Ha} is not an integer. We
will no consider the pseudo ratchet solutions which appear when
\begin{equation} g = 1 + 1/N
\end{equation} where $N$ is an integer. Since $\sin(\omega (1+1/N)t) =
\sin(\omega t)\cos(\omega/N~t)+\cos(\omega t)\sin(\omega/N~t)$ we can
expect to see some effect of some convolution with the longer period
$N/\omega$, but what is surprising is that the coherent dynamics is
intact. We then get the pseudo ratchet solutions of
Fig.~\ref{fig:pseudo-ratchet} for $N=2, 3, 4$ and 5 in which the
ratchet changes direction after $N$ periods.



\section{Summary}
We have done theoretical modeling of the dynamics of small
paramagnetic beads manipulated by the stripe domain pattern of a
ferrite garnet film. The simple model, only taking into account the
paramagnetic interaction and linear hydrodynamics, is sufficient to
explain several non-trivial dynamical phenomenon occurring in the
system. In particular the model is capable of explaining the rectified
motion occurring in a harmonic external field and also the more
complicated separation effect, in which a driving with different
frequencies in $x$~and $z$~direction cause beads of different heights
to go in different directions. We find that this effect can be
explained as a consequence of an asymmetry in the magnetic interaction
potential.


\begin{thebibliography}{6}%
\makeatletter
\providecommand \@ifxundefined [1]{%
 \@ifx{#1\undefined}
}%
\providecommand \@ifnum [1]{%
 \ifnum #1\expandafter \@firstoftwo
 \else \expandafter \@secondoftwo
 \fi
}%
\providecommand \@ifx [1]{%
 \ifx #1\expandafter \@firstoftwo
 \else \expandafter \@secondoftwo
 \fi
}%
\providecommand \natexlab [1]{#1}%
\providecommand \enquote  [1]{``#1''}%
\providecommand \bibnamefont  [1]{#1}%
\providecommand \bibfnamefont [1]{#1}%
\providecommand \citenamefont [1]{#1}%
\providecommand \href@noop [0]{\@secondoftwo}%
\providecommand \href [0]{\begingroup \@sanitize@url \@href}%
\providecommand \@href[1]{\@@startlink{#1}\@@href}%
\providecommand \@@href[1]{\endgroup#1\@@endlink}%
\providecommand \@sanitize@url [0]{\catcode `\\12\catcode `\$12\catcode
  `\&12\catcode `\#12\catcode `\^12\catcode `\_12\catcode `\%12\relax}%
\providecommand \@@startlink[1]{}%
\providecommand \@@endlink[0]{}%
\providecommand \url  [0]{\begingroup\@sanitize@url \@url }%
\providecommand \@url [1]{\endgroup\@href {#1}{\urlprefix }}%
\providecommand \urlprefix  [0]{URL }%
\providecommand \Eprint [0]{\href }%
\providecommand \doibase [0]{http://dx.doi.org/}%
\providecommand \selectlanguage [0]{\@gobble}%
\providecommand \bibinfo  [0]{\@secondoftwo}%
\providecommand \bibfield  [0]{\@secondoftwo}%
\providecommand \translation [1]{[#1]}%
\providecommand \BibitemOpen [0]{}%
\providecommand \bibitemStop [0]{}%
\providecommand \bibitemNoStop [0]{.\EOS\space}%
\providecommand \EOS [0]{\spacefactor3000\relax}%
\providecommand \BibitemShut  [1]{\csname bibitem#1\endcsname}%
\let\auto@bib@innerbib\@empty
\bibitem [{\citenamefont {Vieira}\ \emph {et~al.}(2009)\citenamefont {Vieira},
  \citenamefont {Henighan}, \citenamefont {Chen}, \citenamefont {Hauser},
  \citenamefont {Yang}, \citenamefont {Chalmers},\ and\ \citenamefont
  {Sooryakumar}}]{vieira09}%
  \BibitemOpen
  \bibfield  {author} {\bibinfo {author} {\bibfnamefont {G.}~\bibnamefont
  {Vieira}}, \bibinfo {author} {\bibfnamefont {T.}~\bibnamefont {Henighan}},
  \bibinfo {author} {\bibfnamefont {A.}~\bibnamefont {Chen}}, \bibinfo {author}
  {\bibfnamefont {A.~J.}\ \bibnamefont {Hauser}}, \bibinfo {author}
  {\bibfnamefont {F.~Y.}\ \bibnamefont {Yang}}, \bibinfo {author}
  {\bibfnamefont {J.~J.}\ \bibnamefont {Chalmers}}, \ and\ \bibinfo {author}
  {\bibfnamefont {R.}~\bibnamefont {Sooryakumar}},\ }\href@noop {} {\bibfield
  {journal} {\bibinfo  {journal} {Phys. Rev. Lett.}\ }\textbf {\bibinfo
  {volume} {103}},\ \bibinfo {pages} {128101} (\bibinfo {year}
  {2009})}\BibitemShut {NoStop}%
\bibitem [{\citenamefont {Smistrup}\ \emph {et~al.}(2007)\citenamefont
  {Smistrup}, \citenamefont {Bruus},\ and\ \citenamefont
  {Hansen}}]{smistrup07}%
  \BibitemOpen
  \bibfield  {author} {\bibinfo {author} {\bibfnamefont {K.}~\bibnamefont
  {Smistrup}}, \bibinfo {author} {\bibfnamefont {H.}~\bibnamefont {Bruus}}, \
  and\ \bibinfo {author} {\bibfnamefont {M.~F.}\ \bibnamefont {Hansen}},\
  }\href@noop {} {\bibfield  {journal} {\bibinfo  {journal} {J. Mag. Magn.
  Mat}\ }\textbf {\bibinfo {volume} {311}},\ \bibinfo {pages} {409} (\bibinfo
  {year} {2007})}\BibitemShut {NoStop}%
\bibitem [{\citenamefont {Tierno}\ \emph {et~al.}(2009)\citenamefont {Tierno},
  \citenamefont {Johansen},\ and\ \citenamefont {Fischer}}]{tierno09}%
  \BibitemOpen
  \bibfield  {author} {\bibinfo {author} {\bibfnamefont {P.}~\bibnamefont
  {Tierno}}, \bibinfo {author} {\bibfnamefont {F.~S. T.~H.}\ \bibnamefont
  {Johansen}}, \ and\ \bibinfo {author} {\bibfnamefont {T.~M.}\ \bibnamefont
  {Fischer}},\ }\href@noop {} {\bibfield  {journal} {\bibinfo  {journal} {Phys.
  Chem. Chem. Phys.}\ }\textbf {\bibinfo {volume} {11}},\ \bibinfo {pages}
  {9615} (\bibinfo {year} {2009})}\BibitemShut {NoStop}%
\bibitem [{\citenamefont {Tierno}\ \emph {et~al.}(2007)\citenamefont {Tierno},
  \citenamefont {Reddy}, \citenamefont {Johansen},\ and\ \citenamefont
  {Fisher}}]{tierno07}%
  \BibitemOpen
  \bibfield  {author} {\bibinfo {author} {\bibfnamefont {P.}~\bibnamefont
  {Tierno}}, \bibinfo {author} {\bibfnamefont {S.~V.}\ \bibnamefont {Reddy}},
  \bibinfo {author} {\bibfnamefont {T.~H.}\ \bibnamefont {Johansen}}, \ and\
  \bibinfo {author} {\bibfnamefont {T.~M.}\ \bibnamefont {Fisher}},\
  }\href@noop {} {\bibfield  {journal} {\bibinfo  {journal} {Phys. Rev. E}\
  }\textbf {\bibinfo {volume} {75}},\ \bibinfo {pages} {041404} (\bibinfo
  {year} {2007})}\BibitemShut {NoStop}%
\bibitem [{\citenamefont {Tierno}\ \emph {et~al.}(2008)\citenamefont {Tierno},
  \citenamefont {Reddy}, \citenamefont {Roper}, \citenamefont {Johansen},\ and\
  \citenamefont {Fisher}}]{tierno08}%
  \BibitemOpen
  \bibfield  {author} {\bibinfo {author} {\bibfnamefont {P.}~\bibnamefont
  {Tierno}}, \bibinfo {author} {\bibfnamefont {S.~V.}\ \bibnamefont {Reddy}},
  \bibinfo {author} {\bibfnamefont {M.~G.}\ \bibnamefont {Roper}}, \bibinfo
  {author} {\bibfnamefont {T.~H.}\ \bibnamefont {Johansen}}, \ and\ \bibinfo
  {author} {\bibfnamefont {T.~M.}\ \bibnamefont {Fisher}},\ }\href@noop {}
  {\bibfield  {journal} {\bibinfo  {journal} {J. Phys. Chem. B}\ }\textbf
  {\bibinfo {volume} {112}},\ \bibinfo {pages} {3833} (\bibinfo {year}
  {2008})}\BibitemShut {NoStop}%
\bibitem [{\citenamefont {Helseth}\ \emph {et~al.}(2006)\citenamefont
  {Helseth}, \citenamefont {Wen},\ and\ \citenamefont {Fischer}}]{helseth06}%
  \BibitemOpen
  \bibfield  {author} {\bibinfo {author} {\bibfnamefont {L.~E.}\ \bibnamefont
  {Helseth}}, \bibinfo {author} {\bibfnamefont {H.~Z.}\ \bibnamefont {Wen}}, \
  and\ \bibinfo {author} {\bibfnamefont {T.~M.}\ \bibnamefont {Fischer}},\
  }\href@noop {} {\bibfield  {journal} {\bibinfo  {journal} {J. Appl. Phys.}\
  }\textbf {\bibinfo {volume} {99}},\ \bibinfo {pages} {024909} (\bibinfo
  {year} {2006})}\BibitemShut {NoStop}%
\end{thebibliography}
\end{document}